\documentclass[aps,prl,twocolumn,10pt,superscriptaddress]{revtex4-2}

\pdfoutput=1
\usepackage{lipsum}
\usepackage[dvipsnames]{xcolor}
\usepackage{graphicx,epic,eepic,epsfig}
\usepackage[export]{adjustbox}
\usepackage{tikz,contour} 
\usepackage{feynmf}
\usetikzlibrary{shapes,arrows,positioning,automata,backgrounds,calc,er,patterns}
\usepackage[compat=1.1.0]{tikz-feynman}
\usepackage{tikz}
\usepackage{amsmath}
\usepackage{graphicx}
\usepackage{fullpage}
\usepackage{relsize}
\usepackage[left=1.8cm,right=1.8cm,top=1.8cm,bottom=1.8cm]{geometry}
\usepackage{XCharter}
\usepackage{mathptmx}
\usepackage[T1]{fontenc}
\usepackage{pifont}
\usepackage{booktabs,amsfonts,amsmath,amssymb,amsbsy,bm,mathtools,slashed,latexsym,esvect,simpler-wick}

\usepackage{float,multirow,array,cancel,dcolumn}
\usepackage{enumitem}
\usepackage{hyperref}
\hypersetup{colorlinks=true,citecolor=red,linkcolor=NavyBlue,urlcolor=NavyBlue}
\usepackage[caption=false,labelformat=simple]{subfig}
\usepackage{lipsum}

\usepackage{comment}
\usepackage{shorthand}
\usepackage{natbib}

\let\oldsection\section
\renewcommand{\section}[1]{%
  \par\vspace{1ex} 
  \noindent\textit{\bfseries #1}.--- 
  \ignorespaces
}

\begin{document}
\relscale{1.05}

\title{Photon-initiated enhancements in the pair production of highly charged coloured particles}

\author{Tanumoy Mandal}
\email{tanumoy@iisertvm.ac.in}
\affiliation{Indian Institute of Science Education and Research Thiruvananthapuram, Vithura, Kerala, 695 551, India}

\author{Subhadip Mitra}
\email{subhadip.mitra@iiit.ac.in}
\affiliation{Center for Computational Natural Sciences and Bioinformatics, International Institute of Information Technology, Hyderabad 500 032, India}
\affiliation{Center for Quantum Science and Technology, International Institute of Information Technology, Hyderabad 500 032, India}

\author{Rachit Sharma}
\email{rachit21@iisertvm.ac.in}
\affiliation{Indian Institute of Science Education and Research Thiruvananthapuram, Vithura, Kerala, 695 551, India}

\begin{abstract}\noindent
Strong interaction is typically assumed to dominate the pair production of heavy coloured resonances at the LHC. However, mixed QCD-QED contributions from gluon-photon ($g\gamma$) initial states become critical for highly charged states. This contribution scales with the square of their electric charges and maximises for particles in the fundamental colour representation. We study this effect for leptoquarks, which are colour-triplet bosons. We demonstrate that tree-level mixed QCD-QED contributions enhance their pair-production rates by up to $\sim 33\%$ for a charge-$5/3$ state, rivalling the size of next-to-leading-order QCD corrections. The asymmetric colour flow of $g\gamma$ fusion affects the radiation pattern, altering jet multiplicities and angular distributions. By recasting the latest ATLAS $\mu\mu jj$ search data, we find that these often-overlooked QED effects systematically strengthen mass exclusion limits, establishing a necessary precision standard for bounding highly charged coloured states.
\end{abstract}

\maketitle 


\section{Introduction}
Direct searches for TeV-scale coloured particles at the Large Hadron Collider (LHC) typically focus on their pair production (PP). Because these processes are strong-interaction dominated, the QCD cross sections are used to draw mass exclusions that are insensitive to unknown couplings to the Standard Model (SM) particles. However, if these new states also carry a non-zero electric charge, photon-mediated electromagnetic interactions contribute to the PP processes as well. They also receive contributions from both strong and electromagnetic (mixed QCD-QED) interactions from gluon-photon ($g\gamma$) initial states. Though the photon density in a proton is much smaller than that of the gluon, these contributions can become phenomenologically relevant for highly charged particles, modifying both the total production rates and the inferred mass limits.

For a fixed electric charge, the relative importance of the $g\gamma$-initiated PP channel strictly decreases as the dimension of the colour representation increases, owing to the more rapid growth of the pure QCD-mediated PP cross section. Consequently, the importance of the mixed QCD-QED is most pronounced for colour-triplet states, making leptoquarks (LQs) uniquely suited to demonstrate these production dynamics. LQs arise in a wide class of SM extensions, such as grand unified theories and composite frameworks~\cite{Pati:1974yy,Georgi:1974sy,Fritzsch:1974nn,Farhi:1980xs,Schrempp:1984nj,Wudka:1985ef,Barbier:2004ez}; they are naturally predicted at the TeV scale~\cite{Dorsner:2016wpm}. Both bottom-up and top-down constructions have received renewed attention in light of flavour and precision anomalies, particularly as explanations for the muon anomalous magnetic moment~\cite{Athron:2021iuf}. Consequently, their collider phenomenology has been extensively explored (see, e.g., Refs.~\cite{Bhaskar:2023ftn,Das:2025osr} for the latest collider limits). While higher-order QCD effects are considered in the experimental searches as corrections to the leading order (LO) cross section, the tree-level $g\gamma$-initiated contribution can sizeably enhance the total PP rate, particularly for LQs carrying large electric charges.

In this letter, we systematically quantify the impact of mixed QCD-QED contributions on the PP rates of particles with varying spins and colour representations. Having established the case for LQs,  we recast the latest ATLAS $\mu\mu jj$ search~\cite{ATLAS:2020dsk} to derive updated mass limits for all scalar and vector LQ species, consistently incorporating the tree-level $g\gamma$ contribution. Furthermore, for scalar LQs, we compute PP cross sections at next-to-leading order (NLO) in QCD matched to the \textsc{Pythia} parton shower. We identify certain kinematic observables, such as jet multiplicity, that exhibit characteristic differences between pure QCD and $g\gamma$-initiated channels.

\section{Colour scaling in mixed QCD-QED production}
We first examine how coloured particles can be pair-produced from a $g\gamma$ initial state. Unlike pure-QCD $gg$ fusion, which populates multiple colour configurations and involves complex interference, the $g\gamma$ channel is uniquely constrained. Because the photon is a colour singlet and the gluon is a colour octet, the final $X_R \overline{X}_R$ system must be produced in a colour-octet configuration. Here, $X_R$ denotes a generic coloured particle transforming under a non-trivial irreducible representation $R$ of the $SU(3)_c$. Since the tensor product $R \otimes \overline{R}$ always contains the adjoint representation of $SU(3)$, $g\gamma$-initiated PP is generically allowed for any coloured particle.

For the following process of our concern,
\begin{align}
g + \gamma \to X_R + \overline{X}_R,    
\end{align}
the gluon-$X_R$ coupling, proportional to $(T_R^a)_{ij}$, determines the colour structure of the amplitude. The $g\gamma$ cross section therefore scales linearly with the Dynkin index $T(R)$:
\begin{equation}
\sigma^{g\gamma}_R \sim \sum_{a}\textrm{Tr}(T_R^aT_R^a) = (N_c^2-1)T(R).
\end{equation}
This index is linked to the dimension of the representation $d(R)$ and quadratic Casimir $C_2(R)$ via $8\,T(R) = d(R)C_2(R)$. The number of colour is denoted by $N_c$, which is $3$ here. Expressed in terms of the Dynkin labels $(p,q)$, this evaluates to 
\begin{align}
T(p,q) =&\ \frac{1}{48}(p+1)(q+1)(p+q+2)\nonumber\\
&\  \times\  (p^2 + q^2 + pq + 3p + 3q).    
\end{align}
Calculating this for the fundamental triplet $\mathbf{3}\,(1,0)$, sextet $\mathbf{6}\,(2,0)$, and adjoint octet $\mathbf{8}\,(1,1)$ yields $1/2$, $5/2$, and $3$, respectively. For a fixed particle mass and spin, this directly leads to the following ratios:
\begin{align}
\sigma^{g\gamma}_{\mathbf{3}}:\sigma^{g\gamma}_{\mathbf{6}}:\sigma^{g\gamma}_{\mathbf{8}} = 1:5:6.    
\end{align}
The particle $X_R$ can also be produced through $q\bar{q}$ and $gg$-initiated processes. The cross sections for the $q+\bar{q}\to X_R+\overline{X}_R$ channel follow the same ratios as obtained for $g\gamma$,
\begin{align}
\sigma^{q\bar{q}}_{\mathbf{3}}:\sigma^{q\bar{q}}_{\mathbf{6}}:\sigma^{q\bar{q}}_{\mathbf{8}} = 1:5:6,
\end{align}
since the $q\bar{q}$ channel proceeds through a gluon-mediated $s$-channel diagram. However, the dominant $gg \to X_R+\overline{X}_R$ channel does not follow these simple ratios. The pure QCD cross section grows much more aggressively for higher-dimensional representations due to multiple vertices, non-Abelian gluon self-interactions, and higher-order Casimirs. (We outline some details of these calculations in the Appendix.)

To quantify the relative importance of the $g\gamma$-initiated contribution, we define the ratio
\begin{align}
\mathcal{R} = \dfrac{\sigma^{g\gamma}}{\sigma^{gg} + \sigma^{q\bar{q}}} = \dfrac{\sigma_{\alpha_s\alpha_e}}{\sigma_{\alpha_s^2}}.\label{eq:ratio}
\end{align}
Here, $\alpha_s$ and $\alpha_e$ are the QCD coupling constant and fine-structure constant, respectively. 
Because the pure QCD cross section outpaces the linear $g\gamma$ scaling, this ratio $\mathcal{R}$ is inherently maximised for the lowest-dimensional coloured states. From Table~\ref{tab:pp_cross_sections}, we observe that for $M_{X_R} = 2$~TeV, the ratio $\mathcal{R}$ is significantly larger for a unit-charged triplet, reaching about $12\%$, compared to higher colour representations. Furthermore, the spin of $X_R$ introduces distinct dynamical differences. A similar trend is observed for vector particles, but in contrast, for fermionic states, $\mathcal{R}$ remains relatively small. If $X_R$ is a scalar or vector, a four-point contact interaction ($g\gamma X_R \overline{X}_R$) contributes alongside the exchange diagrams. For fermions, this contact term is absent, restricting production to exchange diagrams alone and suppressing the $g\gamma$ enhancement.

This enhancement becomes increasingly relevant at higher masses. The underlying reason is the relative behaviour of the PDFs at large momentum fractions: while the gluon PDF falls steeply at large Bjorken-$x$, the photon PDF decreases more mildly, leading to an enhanced relative importance of photon-induced processes in the high-mass regime.

Apart from their cross sections, LQ decays further single them out as optimal candidates. Under the assumption that $X_R^s$ decays into two SM particles, only colour triplet (or anti-triplet) states can mediate two-body decays into quark-lepton pairs, making LQs uniquely responsible for the clean dilepton–dijet signatures arising from renormalisable interactions. A sextet state cannot produce a colour-singlet lepton in its two-body decay, and a similar limitation applies to colour-octet scalars or vectors. A colour-octet fermion carrying lepton number, such as a leptogluon, can decay into a lepton and a gluon through a dimension-5 operator. However, as established above, for fermionic states, the relative importance of the $g\gamma$-initiated contribution to pair production is intrinsically smaller, rendering them less suitable for the effects considered here.

\begin{table*}[t]
\caption{Cross sections of $X_R^s$ pair productions (in fb) via $gg$, $q\bar{q}$, and $g\gamma$-initiated channels at $\sqrt{s}=13$~TeV LHC, assuming unit electric charge ($Q=1$). Here, $X_R^s$ denotes a particle with spin $s$ transforming in an irreducible representation $R$ of $SU(3)$. Results are presented for two benchmark masses, $M_X = 1$ and $2$~TeV. The allowed two-body decay modes of $X_R^s$, consistent with group-theoretic constraints and assuming decays to SM states, are also indicated. The ratio $\mathcal{R}$, defined in Eq.~\eqref{eq:ratio}, is shown for each case. \label{tab:pp_cross_sections}}
{\renewcommand\baselinestretch{1.5}\selectfont\begin{tabular*}{\textwidth}{@{\extracolsep{\fill}}llrrrrrrrr}
\toprule
 & & \multicolumn{4}{c}{$M=1$ TeV} 
 & \multicolumn{4}{c}{$M=2$ TeV} \\
\cmidrule(lr){3-6} \cmidrule(lr){7-10}
$X_{R}^s$ & Decay
 & $\sigma_{gg}$ [fb] &  $\sigma_{q\bar{q}}$ [fb]
 & $\sigma_{g\gamma}$ [fb] 
 & $\mathcal{R}$ (\%)
 &$\sigma_{gg}$ [fb] &  $\sigma_{q\bar{q}}$ [fb]
 & $\sigma_{g\gamma}$ [fb]
 & $\mathcal{R}$ (\%)\\
\midrule

$X^0_{\bf 3}$ & $q\ell$, $\bar{q}\bar{q}$ & $2.5$ & $1.0$ 
  & $0.2$ & $5.7$ & $4.3 \times 10^{-3}$
 & $2.5 \times 10^{-3}$ & $0.8 \times 10^{-3}$ & $11.8$ \\

$X^0_{\bf 6}$ & $qq$ & $51.2$ & $4.8$ 
  & $1.0$ & $1.8$ 
 & $8.8 \times10^{-2}$ & $1.2 \times 10^{-2}$&$3.7 \times 10^{-3}$ & $3.7$ \\

$X^0_{\bf 8}$ & $q\bar{q}$ & $53.7$ & $5.7$ 
  & $1.2$ & $2.0$ 
 & $9.3 \times 10^{-2}$ & $1.5 \times 10^{-2}$&$4.5 \times 10^{-3}$ & $4.2$ \\

\midrule

$X^1_{\bf 3}$ & $q\ell$, $\bar{q}\bar{q}$ & $18.0$ & $10.4$ 
  & $1.5$ & $5.2$ 
 & $2.0 \times 10^{-2}$ & $2.3 \times 10^{-2}$&$3.4 \times 10^{-3}$ & $8.0$ \\

$X^1_{\bf 6}$ & $qq$ & $338.9$ & $51.8$ 
  & $7.4$ & $1.9$ 
 & $3.9\times10^{-1}$ & $1.1\times10^{-1}$ & $1.7\times10^{-2}$ & $3.4$ \\

$X^1_{\bf 8}$ & $q\bar{q}$ & $356.8$ & $62.2$ 
  & $8.9$ & $2.1$ 
 & $4.1\times10^{-1}$ & $1.3\times10^{-1}$&$2.1\times10^{-2}$ & $3.8$ \\

\midrule

$X^{1/2}_{\bf 3}$ & $qV$ & $11.0$ & $16.9$ 
 & $0.9$ & $3.1$ 
 & $1.5 \times 10^{-2}$ & $6.5 \times 10^{-2}$&$2.7 \times 10^{-3}$ & $3.4$ \\

$X^{1/2}_{\bf 6}$ & $qg$ & $217.0$ & $84.5$ 
  & $4.4$ & $1.4$ 
 & $3.1\times10^{-1}$ & $3.2\times10^{-1}$& $1.4\times10^{-2}$ & $2.2$ \\

$X^{1/2}_{\bf 8}$ & $\ell g$& $227.6$ & $101.4$ 
  & $5.3$ & $1.6$ 
 & $3.3\times10^{-1}$ & $3.9\times10^{-1}$ & $1.6\times10^{-2}$ & $2.3$ \\

\bottomrule
\end{tabular*}}
\end{table*}
\begin{table*}[]
\caption{Mass exclusion limits for different LQ species at various orders in $\alpha_s$ and $\alpha_e$. The limits are derived assuming only one coupling is nonzero, leading to a $\mu j$ decay. We assume degenerate masses for all components of a given LQ multiplet. For each LQ state, the corresponding BR to the $\mu j$ mode is indicated in parentheses for the relevant one-coupling scenario. In cases marked with $\star$, the effective BR exceeds $100\%$ because more than one component of the LQ multiplet decays into the $\mu j$ mode, resulting in the $\mu j\mu j$ final state.  \label{tab:SLQYukabcd}}
\centering{\small\renewcommand\baselinestretch{1.5}\selectfont
\begin{tabular*}{\textwidth}{l @{\extracolsep{\fill}}lllllll}
\hline
Model & $\alpha_{s}^{2}$ & $\alpha_{s}^{2} + \alpha_{s}\alpha_{e}$ & $\alpha_{s}^{2}(1 + \alpha_{s})$ & $(\alpha_{s}^{2} + \alpha_{s}\alpha_{e})(1 + \alpha_{s})$ & Model & $\alpha_{s}^{2}$ & $\alpha_{s}^{2} + \alpha_{s}\alpha_{e}$  \\ \hline\hline
$S_1\ (50\%)$ &$1273$ & $1274$ & $1336$ & $1337^{+12}_{-14}$ & $U_1\  (50\%)$ &$1641$ & $1647^{+110}_{-64}$ \\ 
$S_1\  (100\%)$  & $1545$ & $1547$  & $1640$ & $1642^{+41}_{-42}$ & $U_1\  (100\%)$  & $1890$ & $1896^{+108}_{-56}$ \\ 
$\widetilde{S}_1\  (100\%)$ & $1545$ & $1576$  &$1640$ & $1667^{+57}_{-44}$ & $\widetilde{U}_1\  (100\%)$ & $1890$ & $1930^{+100}_{-67}$ \\ 
$R_2\  (200\%)^\star$ & $1687$ &  $1748$ & $1795$ & $1812^{+30}_{-35}$ & $V_2\  (200\%)^\star$ & $2044$ &  $2054^{+58}_{-76}$ \\ 
$R_2\  (100\%)$& $1545$ &   $1593$ & $1640$ & $1682^{+72}_{-47}$ & $V_2\  (100\%)$& $1890$ &   $1915^{+104}_{-62}$ \\ 
$\widetilde{R}_2\  (100\%)$ & $1545$ & $1553$ & $1640$ & $1647^{+43}_{-42}$ & $\widetilde{V}_2\  (100\%)$ & $1890$ & $1892^{+108}_{-55}$ \\ 
$S_3\  (150\%)^\star$ & $1592$  & $1618$  & $1687$ & $1725^{+49}_{-64}$ & $U_3\  (150\%)^\star$ & $1933$  & $2001^{+56}_{-110}$ \\ \hline
\end{tabular*}}
\end{table*}

\section{Canonical leptoquarks and beyond}
We refer to leptoquarks (LQs) that couple exclusively to SM quark-lepton pairs as \emph{canonical} LQs, following the Buchm\"uller-R\"uckl-Wyler classification~\cite{Buchmuller:1986zs}. These consist of five scalar and five vector states. Their quantum numbers follow strictly from the possible contractions of SM fermion bilinears under $SU(2)_L$:
\begin{align}
\mathbf{2}\otimes\mathbf{2} &= \mathbf{1}\oplus\mathbf{3} &(S_1,\widetilde{S}_1,U_1,\widetilde{U}_1,S_3,U_3), \nonumber\\
\mathbf{2}\otimes\mathbf{1} &= \mathbf{2} & (R_2,V_2), \nonumber\\
\mathbf{1}\otimes\mathbf{2} &= \mathbf{2} &(R_2,\widetilde{R}_2,V_2,\widetilde{V}_2).
\end{align}
Because these representations are constructed solely from SM fermions, their maximum electric charge is bounded to $5/3$. The $g\gamma$-initiated diagrams originate from the kinetic terms of these LQs ($\Phi$ for scalars and $\chi$ for vectors):
\begin{align}
\mathcal{L}_{\Phi} &= (D_\mu\Phi)^\dagger(D^\mu\Phi), \nonumber\\
\mathcal{L}_{\chi} &= -\frac{1}{2}(D_\mu\chi_\nu - D_\nu\chi_\mu)^\dagger (D^\mu\chi^\nu - D^\nu\chi^\mu),
\end{align}
where $D_\mu$ is the covariant derivative. For vector LQs, an additional non-minimal renormalisable interaction of the form
\begin{align}
\mathcal{L} = -ig_s(1-\kappa)\,\chi^\dagger_\mu T^a \chi_\nu G^{a\mu\nu},
\end{align}
is allowed, which can significantly affect pure QCD-mediated PP rates. In our calculations, we choose the Yang-Mills choice, $\kappa=1$.

For comparison, the cross sections presented in Table~\ref{tab:pp_cross_sections} assume a unit charge, $Q=1$ for all cases. Because $\sigma^{g\gamma}\propto Q^2$ while the pure QCD contribution is independent of $Q$, the relative importance of photon-initiated processes increases rapidly for higher charges. Constrained by the bilinear contractions above, scalar LQs coupling to SM fermions allow charges up to $Q=5/3$, leading to an enhancement of $\sim 33\%$ at $M=2$~TeV.

For other coloured scalars/vectors decaying to SM particles, the maximal charges are typically smaller (e.g., $Q=4/3$ for $S_6$, $Q=1$ for $S_8$, etc.), resulting in more modest effects. In contrast, if the coloured state couples to Beyond the SM (BSM) fermions, significantly larger charges can arise. As explained in the next paragraph, under certain assumptions, LQs interacting with a SM–BSM fermionic pair allow $Q$ up to $8/3$, while LQs interacting with a BSM–BSM fermionic pair can lead to a maximum $Q$ of $11/3$. Since the enhancement scales as $Q^2$, the total cross section can increase by $\sim 84\%$ ($Q=8/3$) or even $\sim 158\%$ ($Q=11/3$) at $M=2$~TeV.

The maximum allowed charges can be understood in terms of simple gauge-invariant constructions if we assume that the new fermionic multiplets interact with the corresponding SM fermions only through off-diagonal Higgs couplings. In principle, the charge-$8/3$ state can appear in a weak doublet scalar LQ, $(\mathbf{3},\mathbf{2},13/6)$, that can form an invariant Yukawa term when combined with a doublet vectorlike (VL) lepton $(\mathbf{1},\mathbf{2},-3/2)$ and $u_R$. Another possible combination is a weak-triplet scalar,  $(\mathbf{3},\mathbf{3},5/3)$, along with a $(\mathbf{1},\mathbf{3},-1)$ VL lepton and $u_R$. Similarly, a weak quintuplet scalar LQ $(\mathbf{3},\mathbf{5},5/3)$ that couples with a weak-triplet VL lepton $(\mathbf{1},\mathbf{3},-1)$ and VL quark $(\mathbf{3},\mathbf{3},2/3)$ pair simultaneously, it contains a charge-$11/3$ component.

\begin{figure*}[t]
\centering
\includegraphics[width=0.32\textwidth]{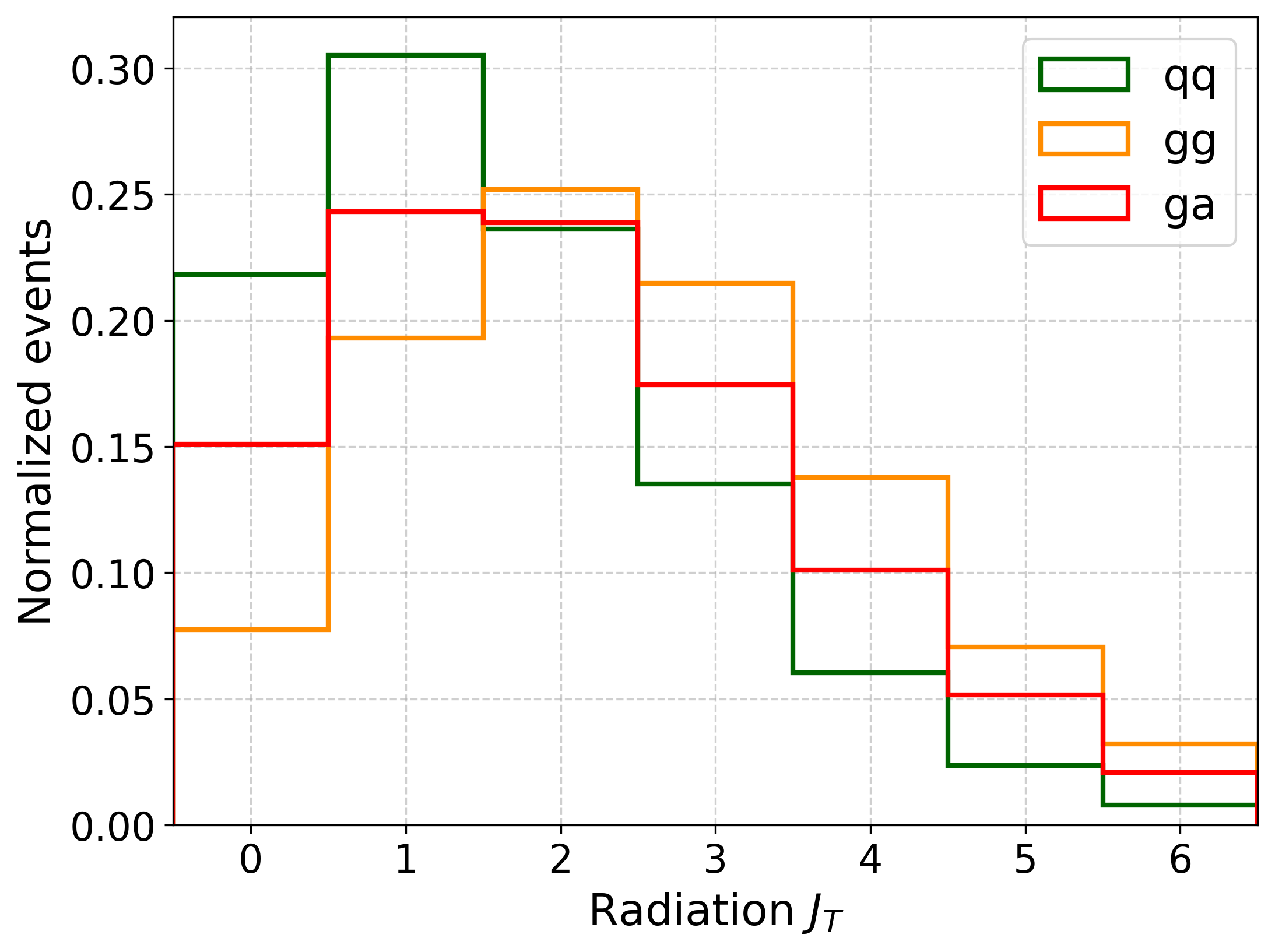}\hfill
\includegraphics[width=0.325\textwidth]{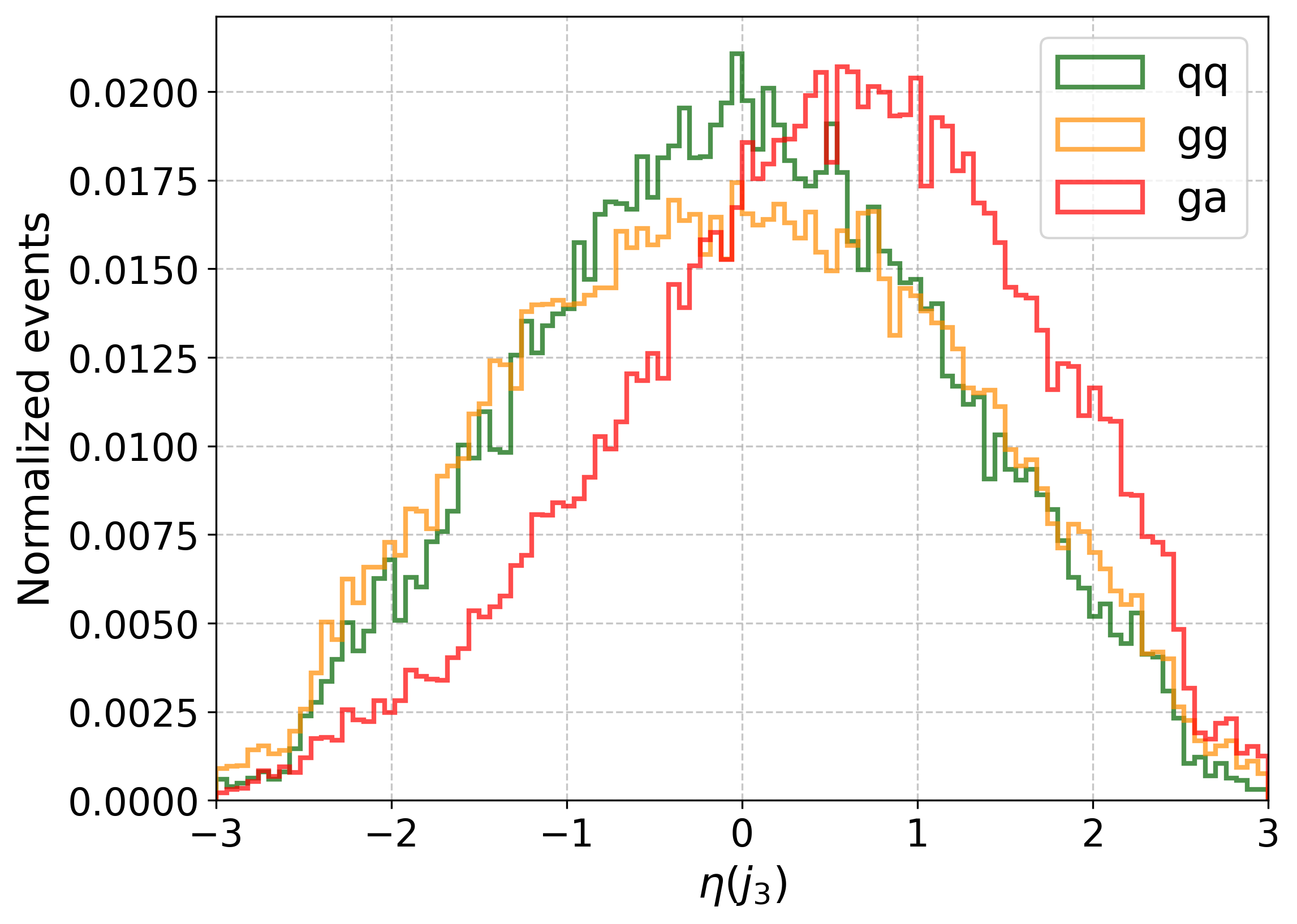}\label{jt3_eta}\hfill
\includegraphics[width=0.325\textwidth]{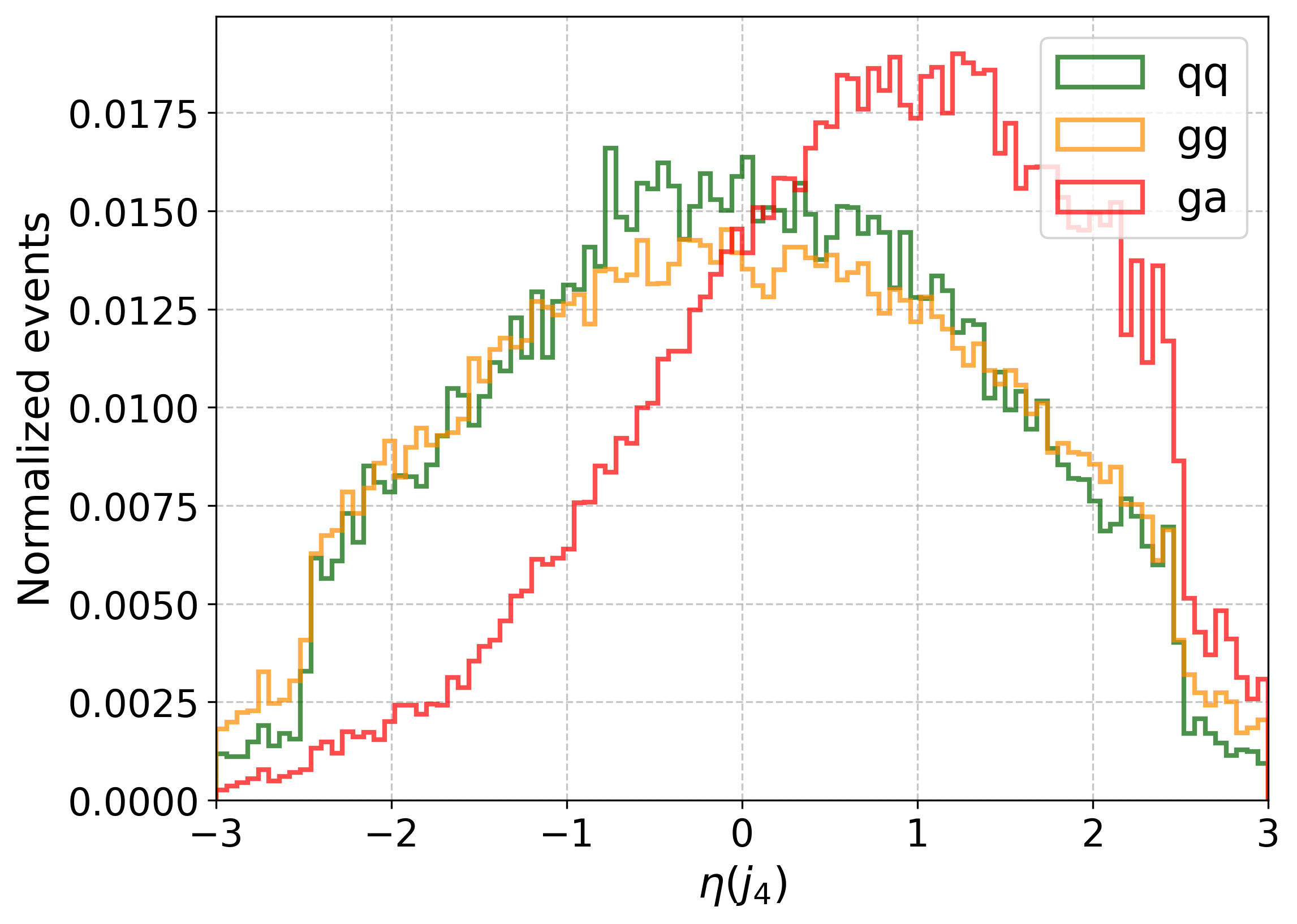}\label{jt4_eta}
\caption{Jet multiplicity distributions (left panel) and pseudorapidity distributions of the hardest (middle panel) and next-to-hardest (right panel) radiation jets for the three subprocesses ($gg$, $q\bar{q}$ and $g\gamma$) contributing to the PP of a scalar LQ of mass $2$~TeV. The radiation off the $g\gamma$ channel is the least central due to the asymmetric colour flow. \label{fig:jetmul}}
\end{figure*}

\section{Effect on current LQ limits} As shown in Table~\ref{tab:pp_cross_sections}, the relative ratio $\mathcal R$ can be sizeable, especially for LQs, which are colour-triplet bosons. Such large contributions should noticeably affect their existing LHC limits. To estimate how the $g\gamma$ channel affects the latest LQ exclusion limits, we implement the relevant interactions in \textsc{FeynRules}~\cite{Alloul:2013bka} to generate \textsc{UFO} model files~\cite{Degrande:2011ua}. Cross sections are evaluated numerically using \textsc{MadGraph5\_aMC@NLO}~\cite{Alwall:2014hca} with a dynamic scale choice, while NLO QCD corrections for scalar LQs~\cite{Mandal:2015lca} are computed via \textsc{NLOCT}~\cite{Degrande:2014vpa}. Scale and PDF uncertainties are estimated using \textsc{SysCalc}~\cite{Kalogeropoulos:2018cke} within \textsc{MadGraph5}. For comparisons and precise estimations, we use different PDF sets from the \textsc{LHAPDF6} database~\cite{Buckley:2014ana}. Because the default \texttt{NNPDF} set in \textsc{MadGraph5} exhibits large uncertainties for the photon density, we consider the \texttt{NNPDFqed}~\cite{Ball:2013hta} and \texttt{LUXqed}~\cite{Butterworth:2015oua,Manohar:2016nzj} PDF sets. However, the former suffers from larger uncertainties in the gluon PDF at high-$x$. Hence, we ultimately employ the \texttt{LUXqed} PDF set for all partons throughout our analysis.

We recast the ATLAS search for the PP of LQs in the $\mu\mu jj$ channel~\cite{ATLAS:2020dsk}. We operate in the limit where the LQ-lepton-quark coupling $\lambda_{\ell q}\to 0$, i.e., where the PP cross section is dominated by gauge interactions, allowing the subleading QED effects to manifest. In this limit, other LQ production modes, such as single or indirect production, give negligible contributions to the $\mu\mu jj$ final state~\cite{Mandal:2015vfa,Mandal:2018kau,Bhaskar:2021pml}. In our calculations, we assume the experimental cut efficiencies are similar across the different PP subprocesses ($gg$, $q\bar{q}$, and $g\gamma$). This is a reasonable assumption as the final-state kinematics are driven by the hard LQ decays, and the inclusive nature of the ATLAS search means that differences in initial-state radiation do not significantly affect signal acceptance. For scalar LQs, we compute cross sections at NLO in QCD~\cite{Mandal:2015lca}, which significantly reduces the scale and PDF uncertainties (combined in quadrature). Conversely, vector LQs are treated strictly at LO, as their NLO predictions are sensitive to the details of the exact nature of ultraviolet completion.

The resulting $\lambda_{\ell_q}$ coupling-independent mass limits are summarised in Table~\ref{tab:SLQYukabcd}. For scalar LQs, we present limits at LO QCD $[\alpha_s^2]$, LO QCD$+$QED $[\alpha_s^2 + \alpha_s\alpha_e]$, NLO QCD $[\alpha_s^2(1 + \alpha_s)]$, and NLO QCD$+$QED $[(\alpha_s^2 + \alpha_s\alpha_e)(1 + \alpha_s)]$ accuracy, whereas vector LQs are shown only at LO QCD and LO QCD$+$QED. Note that in all cases, we have ignored negligibly small pure QED-mediated contributions to PP. Clearly, the mixed QCD-QED contributions have a significant impact on LQs carrying a large electric charge. The limits for the $R_2$ and $U_3$ models, which contain states with $Q=5/3$, show the most pronounced shifts relative to the pure QCD predictions. Furthermore, the inclusion of NLO QCD corrections noticeably strengthens the robustness and stringency of the exclusion limits for scalar LQs.

\section{Effect on kinematic distributions}
Beyond increasing the total PP cross section, mixed QCD-QED processes have a different kinematics than pure QCD ones. Because initial-state radiation differs across partonic channels, observables such as the jet-multiplicity distribution exhibit characteristic differences between the $gg$, $q\bar{q}$, and $g\gamma$ initial states (as illustrated in Fig.~\ref{fig:jetmul} for a scalar LQ of mass $2$~TeV). This is a consequence of the asymmetric colour flow: in $gg$ and $q\bar{q}$ production, both incoming partons carry colour charge, creating a symmetric dipole that copiously radiates gluons, whereas only one initial leg radiates via strong interaction in the $g\gamma$ channel. This suppresses the initial-state radiation profile, resulting in cleaner events with fewer jets.

For scalar and vector LQs, the $g\gamma$ channel includes the $g\gamma X_R \overline{X}_R$ four-point contact interaction, which lacks the $1/\hat{s}$ suppression of $s$-channel exchange diagrams, producing a characteristically harder $p_{\rm T}$ spectrum for the radiated jets. However, their angular distribution is still dominated by the asymmetric colour flow of the $g\gamma$ initial state. Because only the incoming gluon radiates, the radiation jets are preferentially pulled along the beamline, making the mixed channel relatively less central than the symmetric $gg$ or $q\bar{q}$ mediated ones (see Fig.~\ref{fig:jetmul}). This can be advantageous to experimentally identify $g\gamma$-initiated events. Consequently, the QED enhancement could be even more pronounced in a suitably optimised signal region than in our full phase-space estimates.

\section{Discussion and outlook}
In this letter, we presented coupling-independent mass exclusion limits for all LQ species that decay into SM quarks and leptons. By recasting the latest ATLAS $\mu\mu jj$ analysis, we demonstrated that the mixed photon-gluon-initiated production channel can play a non-negligible role in the high-mass regime. Incorporating the photon contribution can modify and strengthen the extracted mass limits, particularly for LQs carrying large electric charges.

The necessity of including these mixed QCD-QED contributions becomes clear when compared to higher-order QCD corrections. Similar to LQ, for charge $2/3$ coloured scalar states like squarks, higher-order QCD has been extensively studied; for instance, for stop pair production, the shift from NLO+NLL to NNLO+NNLL is $\sim 4\%$ at $M=2$ TeV~\cite{Beenakker:2016gmf}. In contrast, the mixed QCD-QED effects can enhance the PP cross section of the stop of mass $2$~TeV by up to $\sim 5\%$. For the PP of higher charged and coloured states, the mixed QCD-QED contributions become increasingly important. This establishes that the additional photon-initiated contributions are comparable in size to the higher-order QCD corrections, making them essential for precise theoretical predictions. At partonic centre-of-mass energies well above the electroweak scale, virtual electroweak corrections (Sudakov logarithms) will further affect this tree-level enhancement. 
A complete $\mathcal{O}(\alpha_e)$ calculation incorporating these effects is therefore essential for future precision studies.

The importance of these photon-initiated processes highlights the potential of future $\gamma p$ colliders, such as the $\gamma p$ option of the LHeC or FCC-eh. Operating at centre-of-mass energies in the range of a TeV (LHeC) to $\sim 3$ TeV (FCC-eh), these machines will provide direct access to photon-parton interactions. Unlike the LHC, where photon-initiated contributions enter as subleading corrections to a QCD background, $\gamma p$ colliders will enhance sensitivity to highly charged states. In this environment, production rates scale as $Q^2$ and can be substantially larger than in proton-proton collisions.

Ultimately, as photon PDFs improve and future collider datasets emerge, photon-initiated processes will play an increasingly vital role in constraining a wide class of BSM scenarios featuring electrically charged states. This work serves as a proof of principle, illustrating how accurately accounting for QED-induced contributions is a necessary step towards obtaining reliable, precision-driven constraints on new physics.

\section{Acknowledgements} T.M. acknowledges partial support from the SERB/ANRF, Government of India, through the Core Research Grant No.~CRG/2023/007031.  R.S. acknowledges the Prime Minister's Research Fellowship (PMRF ID: 0802000).

\bibliography{References}
\bibliographystyle{JHEPCust}
\let\section\oldsection
\appendix
\section{Appendix: Colour scaling of the $gg$-initiated production}\label{app:ggcal} 
\noindent
The $gg$-initiated pair production of $X_R$ includes $s$-, $t$-, and $u$-channel exchanges, along with a four-point (seagull) contact interaction exclusive to scalar and vector particles. For the process $g^a + g^b \longrightarrow X_R^i + \overline{X}_R^j$, the total amplitude can be expressed by factoring out the colour structures from the matrix elements as
\begin{align*}
i\mathcal{M} =&\ f^{abc}\left(T_R^c\right)_{ij}\mathcal{M}_s +\left(T_R^aT_R^b\right)_{ij}\mathcal{M}_t\\ 
&\ + \left(T_R^bT_R^a\right)_{ij}\mathcal{M}_u +\left\{T_R^a,T_R^b\right\}_{ij}\mathcal{M}_4\,.
\end{align*}
To isolate the independent colour flows, we decompose the $t$- and $u$-channel colour products into their symmetric and antisymmetric components (i.e., $T_R^a T_R^b = \frac{1}{2}\{T_R^a, T_R^b\} + \frac{1}{2}[T_R^a, T_R^b]$). Grouping these terms allows us to rewrite the amplitude cleanly into a colour-symmetric channel ($\mathcal{S}$) and a colour-antisymmetric channel ($\mathcal{A}$):
\begin{align*}
i\mathcal{M} = \left\{T_R^a, T_R^b\right\}_{ij}\mathcal{S} + \left[T_R^a, T_R^b\right]_{ij}\mathcal{A},    
\end{align*}
where the effective kinematic functions for the symmetric and antisymmetric configurations are explicitly defined as:
\begin{align*}
\mathcal{S} &= \frac{1}{2}(\mathcal{M}_t + \mathcal{M}_u) + \mathcal{M}_4, \\
\mathcal{A} &= -i\mathcal{M}_s + \frac{1}{2}(\mathcal{M}_t - \mathcal{M}_u).
\end{align*}
As a result, the colour-summed squared matrix element neatly decouples into two independent parts:
\begin{align*}
    \sum_{\text{colours}} |\mathcal{M}|^2 = C_{\mathcal S} |\mathcal{S}|^2 + C_{\mathcal A} |\mathcal{A}|^2,
\end{align*}
where $C_{\mathcal S}$ and $C_{\mathcal A}$ are colour factors dictated by the representation $R$ of the particle: $C_{\mathcal S} = \sum_{a,b} \text{Tr} \left( \left\{T_R^a, T_R^b\right\}^2 \right)$ and $C_{\mathcal A} = -\sum_{a,b} \text{Tr} \left( \left[T_R^a, T_R^b\right]^2 \right)$.

\end{document}